\documentclass[aps,pre,showpacs,amsmath,amssymb,amsfonts,twocolumn]{revtex4-1}

\usepackage{graphicx}
\usepackage{dcolumn}
\usepackage{bm}
\usepackage{epsf}
\usepackage{amsmath}
\usepackage{amssymb}
\usepackage{color}

\newcommand{\bra}[1]{\left\langle #1\right|}
\newcommand{\ket}[1]{\left|#1\right\rangle}

\newcommand{\ptr}[2]{\mathrm{tr_{#1}}\left\{#2\right\}}

\newcommand{\pd}{\partial}

\newcommand{\td}{\mathrm{d}}

\newcommand{\e}[1]{\exp{\left(#1\right)}}

\newcommand{\id}{\mathbb{I}}

\newcommand{\bla}{bla\\bla\\bla\\bla\\bla}

\newcommand{\PRA}{Phys. Rev. A}

\newcommand{\PRE}{Phys. Rev. E}
\newcommand{\PRL}{Phys. Rev. Lett.}
\newcommand{\PRX}{Phys. Rev. X}
\newcommand{\EPL}{EPL (Europhys. Lett.)}

\newcommand{\mc}[1]{\mathcal{#1}}

\newcommand{\mrm}[1]{\mathrm{#1}}

\begin{document}

\title{Information driven current in a quantum Maxwell demon} 

\author{Sebastian Deffner}
\email{sebastian.deffner@gmail.com}
\affiliation{Department of Chemistry and Biochemistry and Institute for Physical Science and Technology, University of Maryland, 
College Park, Maryland 20742, USA}
\date{\today}

\begin{abstract}
We describe a minimal model of a quantum Maxwell demon obeying Hamiltonian dynamics. The model is solved exactly, and we analyze its steady-state behavior. We find that writing information to a quantum memory induces a probability current through the demon, which is the quantum analog of the classical Maxwell demon's action. Our model offers a simple and pedagogical paradigm for investigating the thermodynamics of quantum information processing.
\end{abstract}


\maketitle

One of the most prominent statements of the second law of thermodynamics was formulated by Clausius \cite{clausius_64}, namely that 'no process is possible whose sole result is the transfer of heat from a body of lower temperature to a body of higher temperature'. Shortly after this statement had been established the Scottish physicist James Clerk Maxwell proposed a gedankenexperiment exploring its validity \cite{maxwell_1871}. In a hypothetical set-up an intelligent being, \textit{Maxwell's demon}, observes the \textit{microscopic dynamics} of particles of a hot gas coupled to a gas of lower temperature, and uses this \textit{information} to sort the particles according to their velocity. \textit{Macroscopically} this leads to an apparent violation of the second law. Another gedankenexperiment exploring similar ideas was invented by Leo Szil\'{a}rd, in which the gain of information about microscopic processes is used to perform thermodynamic work \cite{szilard_1929}. One of the main lessons to learn from these seminal contributions to the foundations of modern thermodynamics is that not only 'information is physical' \cite{landauer_1991}, but also that processing information is of thermodynamic relevance.

The past few years have seen renewed interest in the thermodynamics of information processing. For instance, contributions were made to the understanding of feedback control in microscopic systems \cite{sag08,sag09,sag10,toy11,sag12,sagawa_2012,abr11,abr12,def12,gra11}, in the experimental study of Landauer's principle \cite{ber12}, and including aspects of quantum information theory \cite{mor11,ved12,kafri_2012}.

Nevertheless, the nature of Maxwell's demon remained somewhat nebulous as its microscopic foundations are beyond the scope of most previous analyses. Although, already Szil\'{a}rd raised the question of how to constructed a purely mechanical analog of the demon \cite{szilard_1929}, only very recently the first examples of minimal and completely solvable models were proposed \cite{man12,mandal_2013,strasberg_2013,barato_2013,barato_2013a}. In particular, the models of \cite{man12,mandal_2013} motivated the notion of an \textit{information reservoir} \cite{deffner_jarzynski_2013}, which represents the first physically rigorous treatment of a mechanical demon's memory.

The purpose of the present paper is the study of a minimal model of a quantum Maxwell demon. Our analysis is motivated by Mandal's and Jarzynski's three state model \cite{man12}, which is a minimal, stochastic, but classical model of Szil\'{a}rd's engine \cite{szilard_1929}. In \cite{man12} the stochastic dynamics are described by a master equation, whose rates are determined by the interaction with a thermal bath.  It turns out that the demon operates as a rectifier of thermal fluctuations, which can be used to perform thermodynamic work. In the present work the model is transferred to the quantum domain and further reduced as we will consider the \textit{quantum demon} to obey Hamiltonian dynamics. Specifically we consider the quantum demon to consist of three energetically degenerated states that interact with a $N$-qubit-stream, the quantum information reservoir. The total quantum system, demon plus information reservoir, evolves under Hamiltonian dynamics, where the total Hamiltonian describes transition rules equivalent to the classical case \cite{man12}. With this model we will address the question whether processing of quantum information can lead to dynamical behavior similar to its classical analog. As a main result we will see that after an initial transient the quantum demon relaxes into a time-periodic stationary state, in which a persistent current is driven through the quantum demon while information is written to the information reservoir. The present analysis constitutes the first example of a minimal model of a quantum Maxwell demon within the Hamiltonian framework established in \cite{deffner_jarzynski_2013}. Thus, the present work is also of pedagogical value as it illustrates the concepts of thermodynamic information processing in a simple manner.

\subparagraph{The model \label{sec:model}}

\begin{figure}
\centering
 \includegraphics[width=0.47\textwidth]{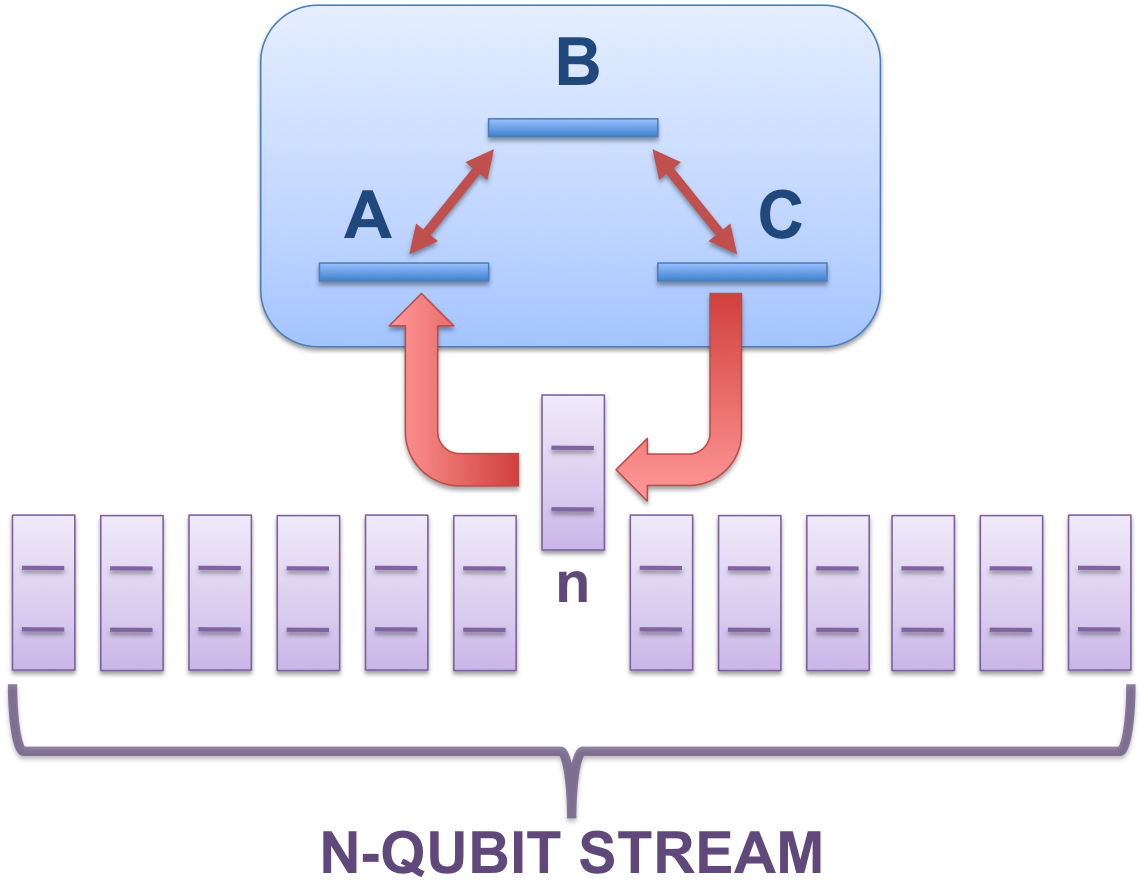}
\caption{\label{qudemon}(color online) Schematic illustration of a quantum demon coupled to an information reservoir realized as a qubit stream. Transitions between states $\ket{A}$ and $\ket{B}$ and between $\ket{B}$ and $\ket{C}$ occur independently of the qubits. Transitions between states $\ket{A}$ and $\ket{C}$ are only possible if at the same time the $n$th qubit is flipped, where the demon exclusively interacts  with the $n$th qubit for times $t\in[(n-1)\tau,\,n\tau]$.}
\end{figure}

Let us start by specifying our model in detail. In complete analogy to the classical case \cite{man12} we consider a quantum system consisting of three energetically degenerated quantum states, $\ket{A}$, $\ket{B}$, and $\ket{C}$, which are coupled to a stream of $N$ qubits, cf. Fig.~\ref{qudemon}. The $N$-qubit stream represents a quantum realization of the \textit{information reservoir} introduced in \cite{deffner_jarzynski_2013}. As for the classical analog \cite{man12} we assume that for times $(n-1) \tau \leq t \leq n \tau$ the demon interacts only with the $n$th qubit. The rules specifying the dynamics are set as follows: We allow for transitions $\ket{A}\leftrightarrow \ket{B}$ and $\ket{B}\leftrightarrow \ket{C}$ without interplay with the qubit stream. Transitions $\ket{C}\rightarrow\ket{A}$ shall only occur if at the same time the $n$th qubit makes a transition from its zero state, $\ket{0}_n$, to its one state, $\ket{1}_n$, and  $\ket{A}\rightarrow\ket{C}$ only when $\ket{1}_n\rightarrow\ket{0}_n$. This set-up is described by the Hamiltonian,
\begin{equation}
\label{eq02}
\begin{split}
&H_\mathrm{tot}(t)=\left(\ket{A}\bra{B}+\ket{B}\bra{A}+\ket{B}\bra{C}+\ket{C}\bra{B}\right)\bigotimes\limits_{n=1}^N \id_n\\
&+ \gamma \sum\limits_{n=1}^N \Pi_n(t)\,\ket{A}\bra{C}\otimes \id_1\otimes \dotsm\otimes \begin{pmatrix}0&1\\0&0\end{pmatrix}_n\otimes\dotsm\otimes \id_N\\
&+ \gamma \sum\limits_{n=1}^N \Pi_n(t)\,\ket{C}\bra{A}\otimes \id_1\otimes \dotsm\otimes \begin{pmatrix}0&0\\1&0\end{pmatrix}_n\otimes\dotsm\otimes \id_N\,,
\end{split}
\end{equation}
where $\Pi_n(t)$ is the Heaviside-$\pi$-function,
\begin{equation}
\label{eq03}
\Pi_n(t)=\begin{cases} 1 & \forall t: (n-1)\,\tau \leq t < n\, \tau\\ 0 & \mathrm{elsewhere}\end{cases}\,,
\end{equation}
and $\id_n$ is the identity operator in the reduced Hilbert space of the $n$th qubit. Finally, by $\gamma$ we denote the coupling strength of the three level quantum demon and the $N$-qubit stream, where we work in units so that $\gamma$ is measured in multiples of $1/\tau$. Note that in the classical model the coupling strength was set to $\gamma=1$ \cite{man12}. However, we will see in the following discussion that the relative value of $\gamma$ determines the overall dynamical behavior of the quantum demon.

It is worth noting that the latter model is not only of academic interest, but can be understood as a real \textit{physical} system. Consider, for instance, a spin-1 particle with Hamiltonian $H=-\mu\,\vec{\sigma}  \circ \vec{B}$, where $\mu$ is the magnetic moment, $\vec{\sigma}$ is the Pauli vector, and $\vec{B}$ denotes the magnetic field. Then $H$ becomes the the reduced Hamiltonian of the demon only, i.e. $H=\ptr{\mathrm{info}}{H_\mathrm{tot}}$, for the specific choice  $\vec{B}=(-\sqrt{2}/\mu, 0, 0)$, where the Pauli matrix $\sigma_x$ is given by \cite{cohen_vol1_77} 
\begin{equation}
\label{eq05}
\sigma_x= \sqrt{2} \begin{pmatrix} 0&1&0 \\ 1&0&1 \\ 0&1&0 \end{pmatrix} \,.
\end{equation}
The $N$-qubit stream can be thought of a set of $N$ non-interacting spin-1/2 particles, at which the spin-1 particle flies by. The additional condition that the quantum demon, here the spin-1 particle, interacts at any instant only with a single qubit can be realized by sufficient spatial separation of the spin-1/2 particles. 

\subparagraph{Solution of the dynamics \label{sec:dynamics}}

In the present analysis we shall be interested in the reduced dynamics of the quantum demon only. The total system is evolving under unitary dynamics,
\begin{equation}
\label{eq06}
i\hbar\,\dot\rho_\mathrm{tot}=\left[H_\mathrm{tot}(t),\rho_\mathrm{tot}\right]
\end{equation}
and the reduced density operator $\rho(t)$ of the quantum demon is obtained by tracing out the information reservoir. We have accordingly,
\begin{equation}
\label{eq07}
\rho(t)=\ptr{\mathrm{info}}{\rho_\mathrm{tot}(t)}=\ptr{\mathrm{info}}{U_\mathrm{tot}(t)\,\rho_\mathrm{tot}(t_0)\,U_\mathrm{tot}^\dagger(t)}\,,
\end{equation}
where $U(t)=\mc{T}_>\,\e{-i/\hbar\,\int_0^t\td s\,H_\mathrm{tot}(s)}$. Note that $\rho(t)$ lives in the $3$-dimensional Hilbert space spanned by $\ket{A}$, $\ket{B}$, and $\ket{C}$. 

In order to analyze the dynamical behavior of the quantum demon we have to solve for the trace preserving completely positive (TCP) map, which determines the time evolution of $\rho(t)$. To this end, we write the TCP map in terms of its Kraus operator expansion \cite{nielsen_00},
\begin{equation}
\label{eq08}
\rho(t)=\sum\limits_\nu  T_\nu\,\rho(t_0)\,T_\nu^\dagger\,.
\end{equation}
In general, to determine the $T_\nu$ is a hardly feasible task. However, the present situation greatly simplifies since the three level system interacts merely with a single qubit at any instant. Specifically, for each time interval $t\in\left[(n-1)\, \tau, n\, \tau\right]$ the total dynamics are effectively determined by the dynamics in the reduced Hilbert space of the demon and the $n$th qubit. Therefore, the TCP map can be constructed from of the following cyclic protocol describing the dynamics of the reduced density operator of the demon only:
\paragraph*{Step 1:} At $t=(n-1)\,\tau$ the combined density operator of the demon and $n$th qubit can be written as the direct product,
\begin{equation}
\label{eq09}
\rho_{\mrm{dem}, n}\left((n-1)\,\tau\right)=\rho\left((n-1)\,\tau\right)\otimes\rho_n(0)\,,
\end{equation}
where we denote the initial state of the $n$th qubit by $\rho_n(0)$. For the latter construction we explicitly used the fact that the demon exclusively interacts with a single qubit at any instant, and the qubits do not interact amongst themselves. If we included, for instance, qubit-quibt interaction Eq.~\eqref{eq09} would not hold due to effects of entanglement. For the sake of simplicity and for pedagogical reasons we leave the study of entanglement in an extended model for future work and focus here on the simplest model.
\paragraph*{Step 2:} For $(n-1)\,\tau\leq t\leq n\,\tau$ demon and $n$th qubit evolve under unitary dynamics generated by the reduced Hamiltonian,
\begin{equation}
\begin{split}
\label{eq10}
H_{\mrm{dem,n}}&=\ket{A0_n}\bra{B0_n}+\ket{B0_n}\bra{A0_n}+\ket{B0_n}\bra{C0_n}\\
&+\ket{C0_n}\bra{B0_n}+\gamma\,\left(\ket{C0_n}\bra{A1_n}+\ket{A1_n}\bra{C0_n}\right)\\
&+\ket{A1_n}\bra{B1_n}+\ket{B1_n}\bra{A1_n}+\ket{B1_n}\bra{C1_n}\\
&+\ket{C1_n}\bra{B1_n}\,,
\end{split}
\end{equation}
which can be written in matrix notation
\begin{equation}
H_{\mrm{dem,n}}=\begin{pmatrix}0&1&0&0&0&0 \\ 1&0&1&0&0&0 \\0&1&0&\gamma&0&0 \\ 0&0&\gamma&0&1&0 \\ 0&0&0&1&0&1 \\ 0&0&0&0&1&0 \end{pmatrix}\,.
\end{equation}
Note that in the quantum case the latter Hamiltonian describes the transition rules given by the rate matrix of the analogous, classical model \cite{man12}.
\paragraph*{Step 3:}
At $t=n\,\tau$ the $n$th qubit is decoupled. Mathematically that means that the $n$th qubit can be traced out, and we have  $\rho\left(n\,\tau\right)=\ptr{\textit{n-}th\, qubit}{\rho_{\mrm{dem}, n}\left(n\,\tau\right)}$. To complete the cycle the protocol starts with \textit{Step 1} again, where the demon now interacts with the $(n+1)$st qubit. 

The Kraus operators $T_\nu$ are then given by a product of operators corresponding to steps \textit{Step 1} to \textit{Step 3} of the latter protocol.

\paragraph*{Time periodic steady state}

All TCP maps have a fixed point \cite{terhal_2000}. If the fixed point is unique the generated time evolution converges towards this fixed point.  In the present case the fixed point can be explicitly evaluated, and hence proven by \textit{calculation} that it is unique. 

The above introduced protocol yields dynamical fixed points for all times during one cycle, $t\in [(n-1)\,\tau, n\,\tau]$ for sufficiently large $n$. This means that the reduced density operator of the demon, $\rho(t)$, relaxes into a time-periodic steady state, where $\rho(n\,\tau)\xrightarrow{n\to\infty}\rho^{\mrm{ss}}$. We continue by illustrating how to construct the Kraus operators $T_\nu$ and the corresponding fixed points of the map by considering specific examples. 

\paragraph*{Explicit example}
For the sake of simplicity let the $N$ qubits all be prepared in the zero state. This represents an initially \textit{blank} memory. We have,
\begin{equation}
\label{eq11}
\rho_n(0)=\begin{pmatrix} 1&0\\ 0&0 \end{pmatrix}_n
\end{equation}
for all $n\in \{1,2,\dots,N\}$.

The first step of the protocol constructs the density operator, $\rho_{\mrm{dem}, n}$, at time $t=(n-1)\,\tau$ as a direct product,
\begin{equation}
\label{eq12}
\rho\left((n-1)\,\tau\right)\otimes\rho_n(0)=B\,\rho\left((n-1)\,\tau\right)\,B^\dagger\,,
\end{equation}
where the operator $B$ reads
\begin{equation}
\label{eq13}
B^\dagger=\begin{pmatrix} 1&0&0&0&0&0 \\ 0&1&0&0&0&0 \\ 0&0&1&0&0&0 \end{pmatrix}\,.
\end{equation}
Note that in the general case $B$ can be much more complicated. In particular, for mixed initial states of the qubits  Eq.~\eqref{eq12} becomes a linear combination, $\rho\left((n-1)\,\tau\right)\otimes\rho_n(0)=\sum_\nu B_\nu\,\rho\left((n-1)\,\tau\right)\,B^\dagger_\nu$.

For times $t\in\left[(n-1)\,\tau,\, n\,\tau\right]$ demon and $n$th qubit evolve under unitary dynamics generated by \eqref{eq10}. For instance, for $\gamma=\gamma_1=4/3$ the eigenenergies, $E$, of $H_{\mrm{dem,n}}$ are given by
\begin{equation}
\label{eq13a}
\begin{split}
E\in\{ &2, -2, 1/3\,(-1-\sqrt{7}), 1/3\,(1-\sqrt{7}),\\
& 1/3\,(-1+\sqrt{7}), 1/3\,(1+\sqrt{7}) \}\,.
\end{split}
\end{equation}
Note that for the sake of simplicity, we work in units were $\tau=1$. With these the propagator $U_{\mrm{dem,n}}=\e{-i/\hbar\, H_{\mrm{dem,n}} t}$ describing the evolution for times $t \in[(n-1)\,\tau, n\,\tau]$ can be evaluated explicitly.

Finally, the partial trace has to be written in operator formulation, as well. We have,
\begin{equation}
\label{eq14}
\begin{split}
\rho\left(n\,\tau\right)&=\ptr{\textit{n-}th\, qubit}{\rho_{\mrm{dem}, n}\left(n\,\tau\right)}\\
&= P_1\, \rho_{\mrm{dem}, n}\left(n\,\tau\right)\,P_1^\dagger+P_2\, \rho_{\mrm{dem}, n}\left(n\,\tau\right)\,P_2^\dagger\,,
\end{split}
\end{equation}
where the operators $P_1$ and $P_2$ are given by
\begin{equation}
\label{eq15}
\begin{split}
P_1&= \begin{pmatrix} 1&0&0&0&0&0 \\ 0&1&0&0&0&0 \\ 0&0&1&0&0&0 \end{pmatrix} \\
P_2&=\begin{pmatrix} 0&0&0&1&0&0 \\ 0&0&0&0&1&0 \\ 0&0&0&0&0&1 \end{pmatrix}\,.
\end{split}
\end{equation}
Note that the latter projection operators are the quantum equivalent of the projection introduced in the solution of the classical model, cf. Ref. \cite{man12}. Accordingly, the time evolution of the reduced density operator, $\rho(t)$, for $t\in\left[(n-1)\,\tau,\, n\,\tau\right]$ is determined by the TCP map,
\begin{equation}
\label{eq16}
\rho(t)=T_1(t)\,\rho\left((n-1)\,\tau\right)\,T_1^\dagger(t)+T_2(t)\,\rho\left((n-1)\,\tau\right)\,T_2^\dagger(t),
\end{equation}
where $T_1(t)=P_1\,U_{\mrm{dem,n}}(t)\,B$ and $T_2(t)=P_2\,U_{\mrm{dem,n}}(t)\,B$. The total time evolution of $\rho(t)$ from initial time $t=0$ to $t\in\left[(n-1)\,\tau,\, n\,\tau\right]$ is constructed as follows: First,  $T_1(\tau)$ and $T_2(\tau)$ are iteratively applied $(n-1)$-times to an initial state $\rho(t=0)$; then, the final state of interest $\rho(t)$ is obtained from Eq.~\eqref{eq16}.

For the particular choice $\gamma_1=4/3$ the fixed point can be calculated analytically. However, the exact expression is still rather complicated. Therefore, we will turn to a numerical analysis in the next section. For comparison of the two approaches, it is instructive, however, to determine a numerical value for the fixed point $\rho^\mrm{fix}_{\gamma_1} $ from the analytical formula. We obtain, that in the beginning of each cycle $\rho(t)$ takes the value 
\begin{equation}
\label{eq17}
\rho^\mrm{fix}_{\gamma_1}=\begin{pmatrix} 0.42 & 0.25 i & -0.09 \\ -0.25 i & 0.35 & 0.22 i \\ -0.09 & -0.22 i & 0.23\end{pmatrix}\,.
\end{equation}
\begin{figure}
\centering
\includegraphics[width=0.48\textwidth]{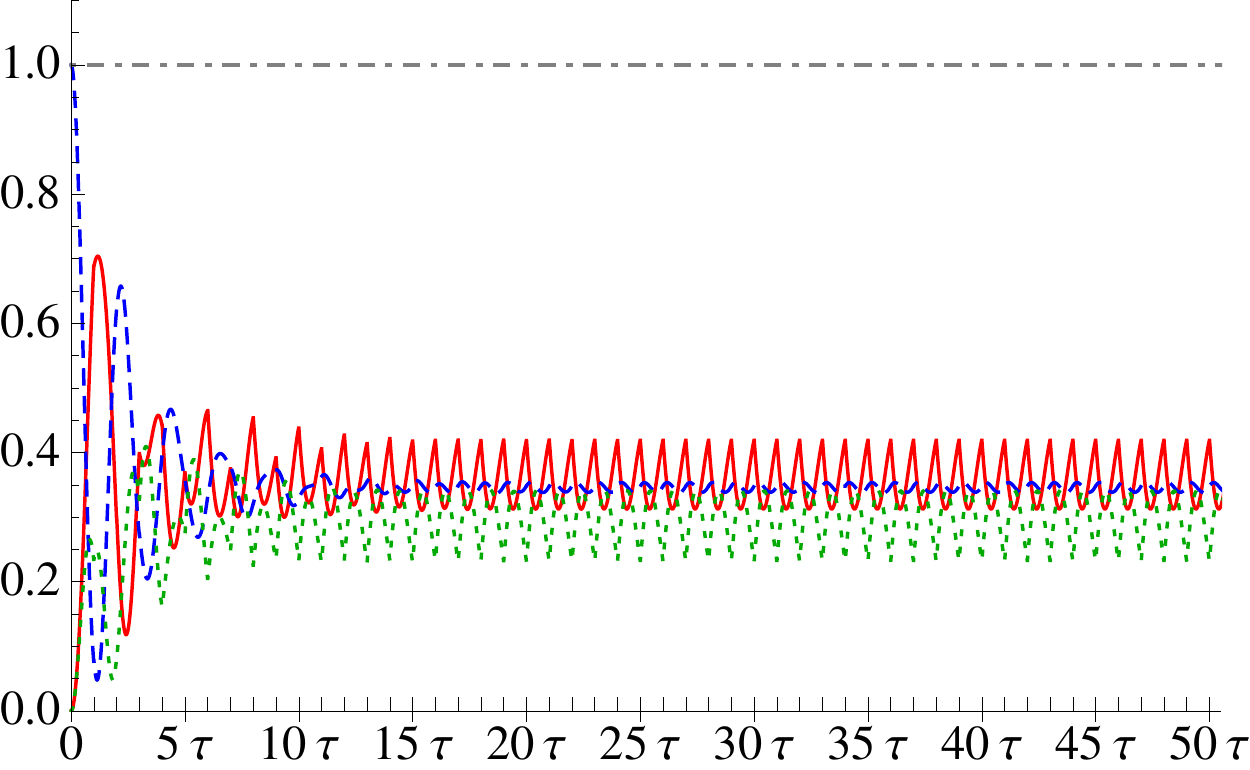}
\caption{\label{fig1} (color online) Occupation probabilities $p_A(t)$ (red, solid line), $p_B(t)$ (blue, dashed line), and $p_C(t)$ (green dotted line) as a function of time for $\gamma=4/3$ and initial state \eqref{eq21}; $\tau$ in arbitrary units.}
\end{figure}
\begin{figure}
\centering
\includegraphics[width=0.48\textwidth]{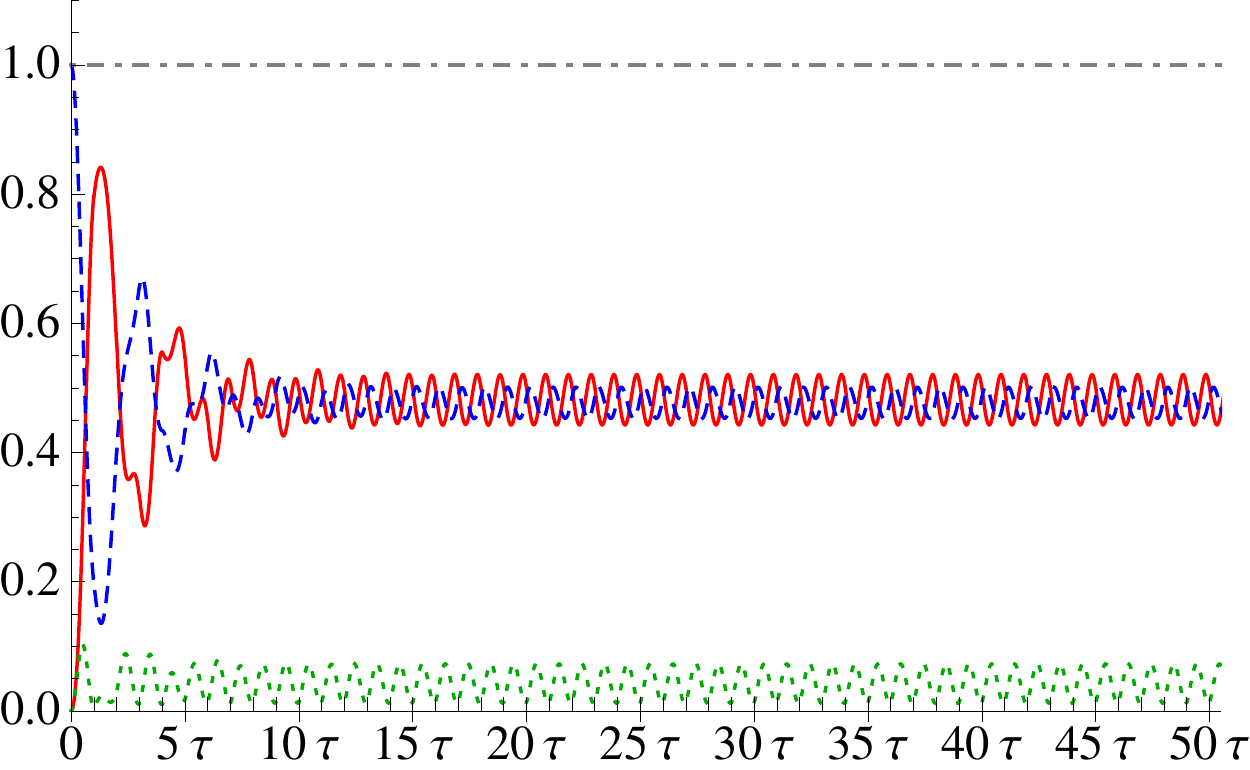}
\caption{\label{fig3}  (color online) Occupation probabilities $p_A(t)$ (red, solid line), $p_B(t)$ (blue, dashed line), and $p_C(t)$ (green dotted line) as a function of time for $\gamma=\gamma_2$ \eqref{eq18} and initial state \eqref{eq21}; $\tau$ in arbitrary units.}
\end{figure}

Another case, where the fixed point can be evaluated analytically is given by 
\begin{equation}
\label{eq18}
\gamma_2=\frac{\pi^3-2 \pi }{\pi^2-1 }\,.
\end{equation}
Then the eigenvalues of $H$ are  
\begin{equation}
\label{eq19}
\begin{split}
E\in&\Bigg\{\pi,\frac{\pi + \sqrt{8 - 11 \pi^2 + 4 \pi^4}}{2 - 2 \pi^2}, \frac{\pi - \sqrt{8 - 11 \pi^2 + 4 \pi^4}}{-2 + 2 \pi^2}, \\
&-\pi, \frac{\pi - \sqrt{8 - 11 \pi^2 + 4 \pi^4}}{2 - 2 \pi^2}, \frac{\pi + \sqrt{8 - 11 \pi^2 + 4 \pi^4}}{-2 + 2 \pi^2}\Bigg\}\,,
\end{split}
\end{equation}
which results in a numerical approximation of the analytical result as,
\begin{equation}
\label{eq20}
\rho^\mrm{fix}_{\gamma_2}=\begin{pmatrix} 0.50 & 0.15 i & -0.07 \\  -0.15 i & 0.49 & 0.05 i \\ -0.07 &  -0.05 i & 0.01\end{pmatrix}\,.
\end{equation}
We observe that the occupation probabilities for states $\ket{A}$, $\ket{B}$, and $\ket{C}$, i.e. the diagonal elements of the reduced density operator for the demon only, are strongly dependent on the value of the coupling strength $\gamma$. 

\paragraph*{Numerical solution}
\begin{figure}
\centering
\includegraphics[width=0.48\textwidth]{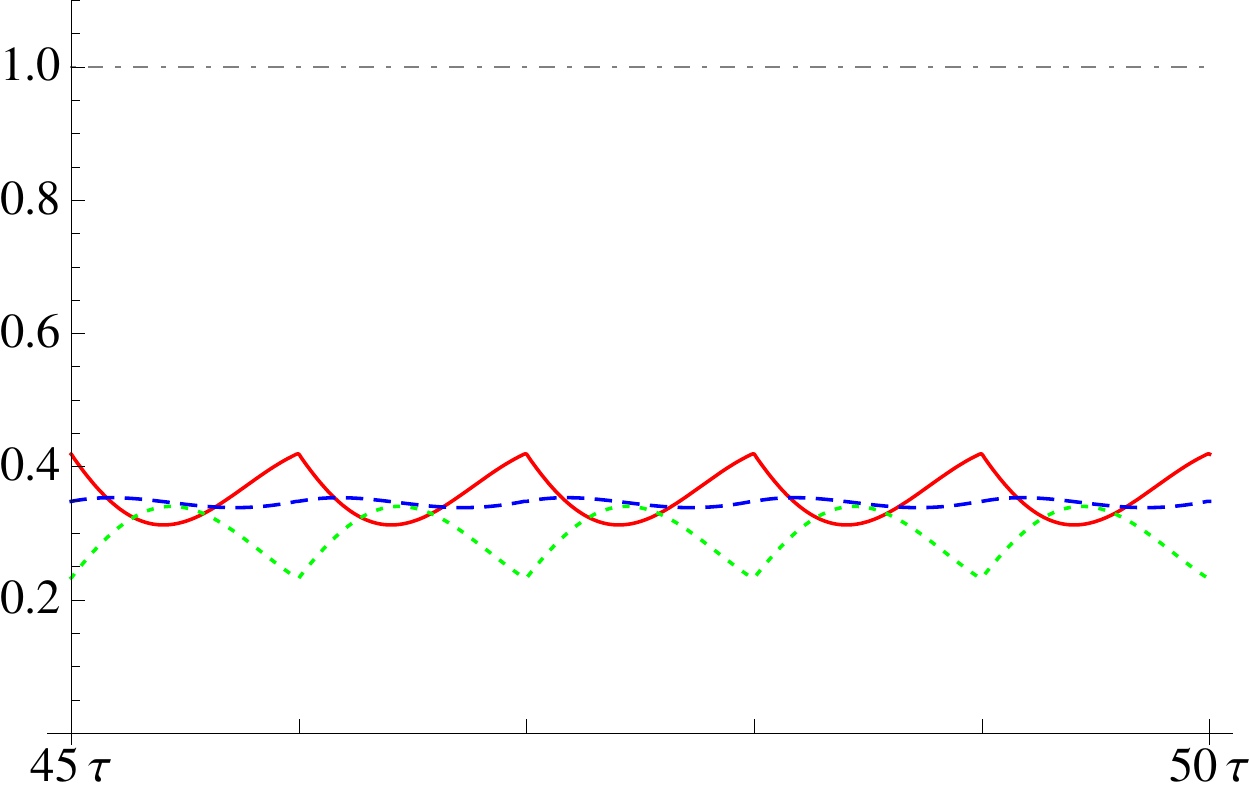}
\caption{\label{fig2} (color online) Occupation probabilities $p_A(t)$ (red, solid line), $p_B(t)$ (blue, dashed line), and $p_C(t)$ (green dotted line) as a function of time in the time periodic stationary state for $\gamma=4/3\tau$; $\tau$ in arbitrary units.}
\end{figure}
\begin{figure}
\centering
\includegraphics[width=0.48\textwidth]{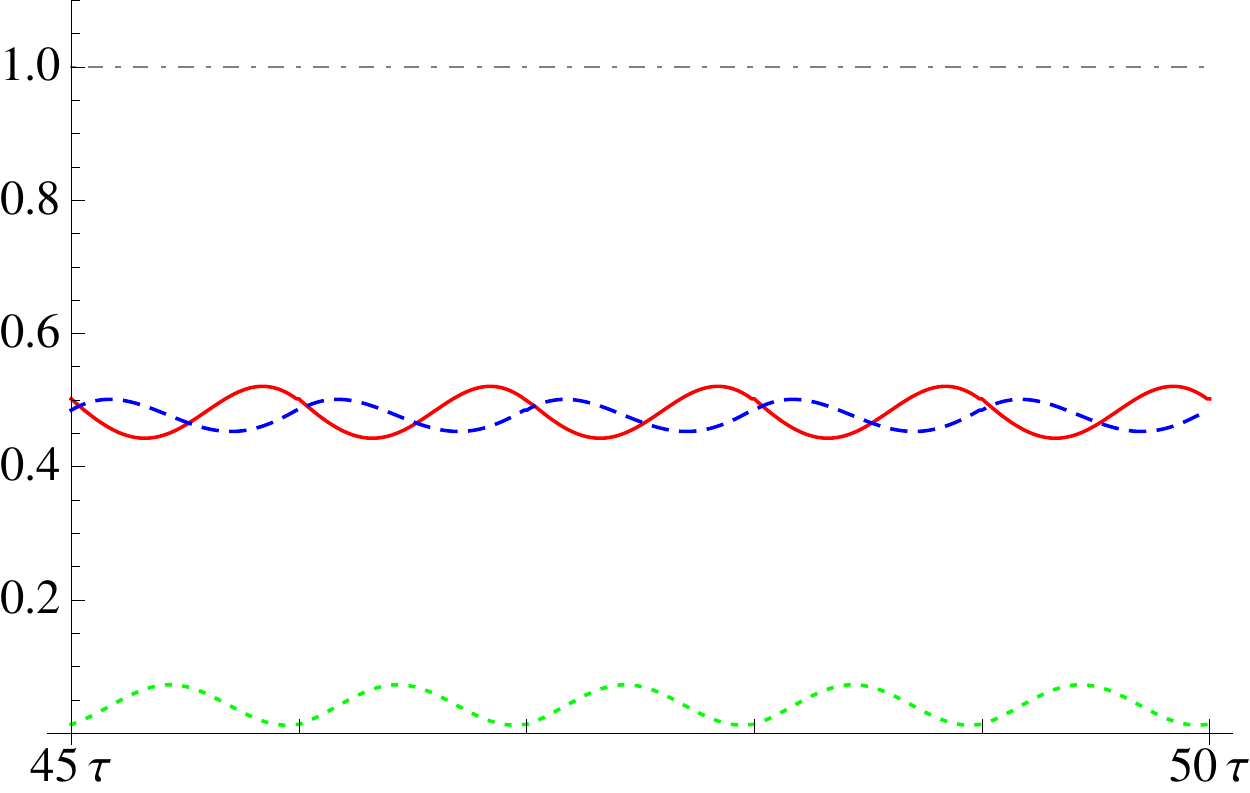}
\caption{\label{fig4}  (color online) Occupation probabilities $p_A(t)$ (red, solid line), $p_B(t)$ (blue, dashed line), and $p_C(t)$ (green dotted line) as a function of time in the time periodic stationary state for $\gamma=\gamma_2$ \eqref{eq18}; $\tau$ in arbitrary units.}
\end{figure}
In order to gain further insight into the dynamics of the quantum demon we also solved numerically for the reduced dynamics. As initial condition we chose,
\begin{equation}
\label{eq21}
\rho(0)=\begin{pmatrix}
0&0&0\\
0&1&0\\
0&0&0
\end{pmatrix}\,,
\end{equation}
in order to exclude any artificial bias potentially arising from the boundary condition. In Figs.~\ref{fig1} and \ref{fig3} we plot the diagonal elements of the demons density operator for $\gamma=\gamma_1=4/3$ and $\gamma=\gamma_2$ \eqref{eq18}. We denote the occupation probabilities as $p_A(t)=\bra{A}\rho(t)\ket{A}$, $p_B(t)=\bra{B}\rho(t)\ket{B}$, and $p_C(t)=\bra{C}\rho(t)\ket{C}$. In both case we observe a significant relaxation into its time-periodic steady state, where in the beginning of each ``qubit interval'' the reduced state takes its fixed point values, which we determined independently and analytically, cf. Eqs.~\eqref{eq17} and \eqref{eq20}. In Figs.~\ref{fig2} and \ref{fig4} we plot $p_A(t)$, $p_B(t)$, and $p_C(t)$ for times for which $\rho(t)$  has relaxed into its time-periodic steady state, here for $n\in\{45,\dots,50\}$, where $n$ is the index of the qubit. We notice that at the beginning of each interacting interval $\rho(t)$ returns to its fixed point values determined in Eqs.~\eqref{eq17} and Eq.~\eqref{eq20}, respectively. The numerical results are in perfect agreement with the prediction of the analytical treatment, i.e. with the explicit calculation of the TCP map describing the time-evolution of the quantum demon.

\subparagraph{Information driven current \label{sec:current}}

\begin{figure}
\centering
\includegraphics[width=0.48\textwidth]{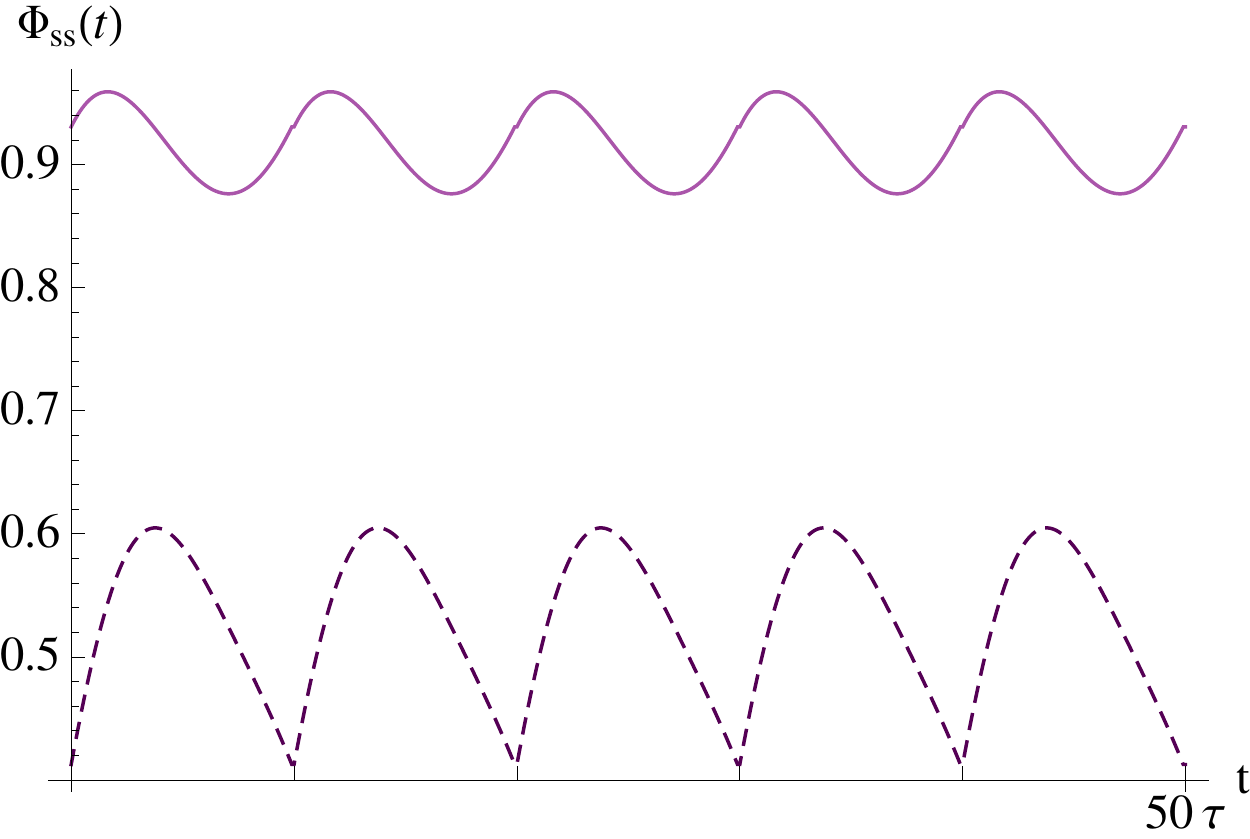}
\caption{\label{phit} (color online) Total current $\Phi_\mrm{ss}(t)$ in the time-periodic stationary state for $\gamma=4/3$ (lighter purple, solid line) and $\gamma=\gamma_2$ \eqref{eq18} (darker purple, dashed line); $\tau$ in arbitrary units.}
\end{figure}

\begin{figure}
\centering
\includegraphics[width=0.48\textwidth]{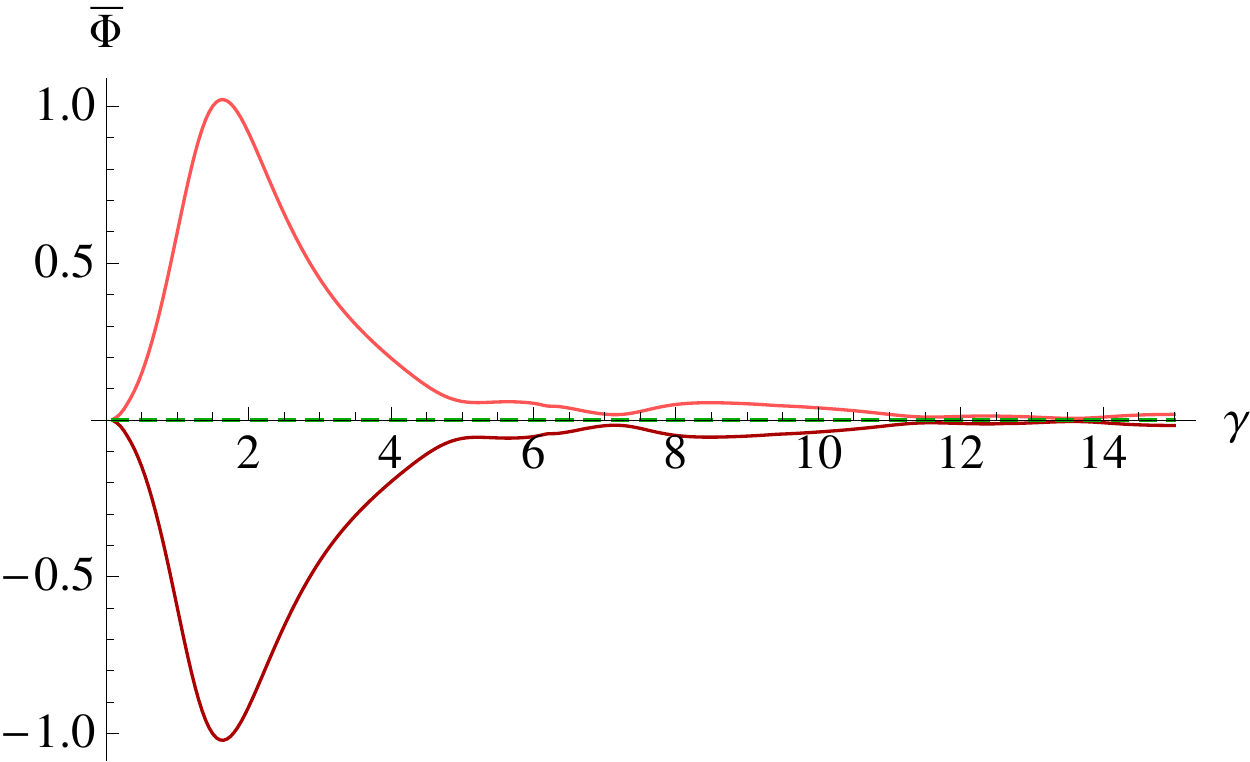}
\caption{\label{phi_gamma} (color online) Average current $\overline{\Phi}$ \eqref{eq26} as a function of $\gamma$ for $\rho_n(0)=\rho_n^1(0)$ (lighter red, upper, solid line), $\rho_n(0)=\rho_n^2(0)$ (darker red, lower, solid line), and $\rho_n(0)=\rho_n^3(0)$ (green, dashed line); $\gamma$ in units $1/\tau$.}
\end{figure}

\begin{figure}
\centering
\includegraphics[width=0.48\textwidth]{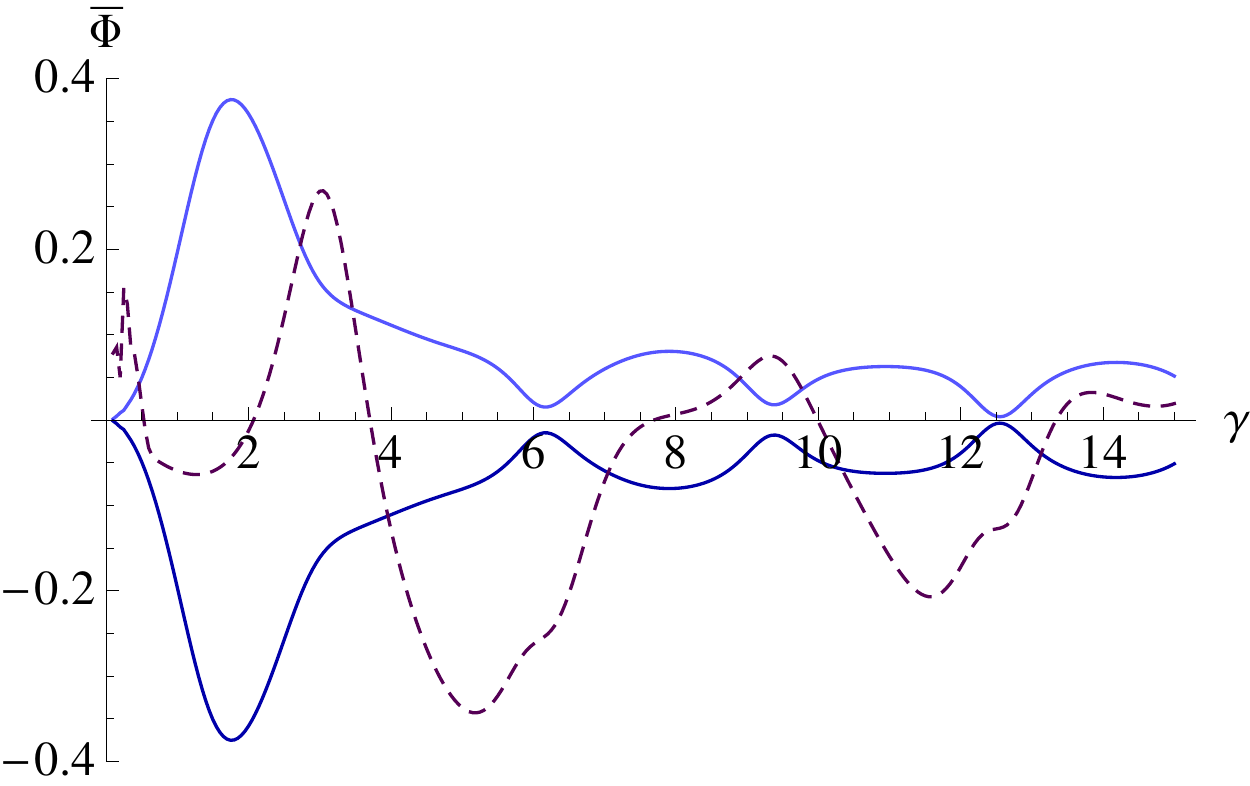}
\caption{\label{phi_gamma2} (color online) Average current $\overline{\Phi}$ \eqref{eq26} as a function of $\gamma$ for $\rho_n(0)=\rho_n^4(0)$ (lighter blue, upper, solid line), $\rho_n(0)=\rho_n^5(0)$ (darker blue, lower, solid line), and $\rho_n(0)=\rho_n^6(0)$ (purple, dashed line); $\gamma$ in units $1/\tau$.}
\end{figure}

The main question is whether the information exchange with information reservoir, here the $N$ qubit stream, can drive a stationary current in the quantum demon. For systems with continuous variables the probability current is defined via its continuity equation \cite{cohen_vol1_77,shankar_94}, 
\begin{equation}
\label{eq22}
\pd_t \bra{x}\rho(t)\ket{x}+\pd_x j(x,t)=0\,.
\end{equation}
It is easy to show \cite{cohen_vol1_77,shankar_94} that the probability current can be written as
\begin{equation}
\label{eq23}
j(x,t)=\frac{i\hbar}{2 m}\,\left[\pd_y \bra{x}\rho(t)\ket{y}-\pd_x \bra{x}\rho(t)\ket{y}\right]\Big|_{x=y}\,,
\end{equation}
which remains valid for open systems dynamics described by Markovian master equations  \cite{breuer_07}. Therefore,  we define, in complete analogy, the probability current in the discrete quantum demon to read 
\begin{equation}
\label{eq24}
\begin{split}
\phi(\nu,t)&\equiv\frac{i\hbar}{2}\left[\bra{\nu-1}\rho(t)\ket{\nu}-\bra{\nu}\rho(t)\ket{\nu-1}\right]\\
&-\frac{i\hbar}{2}\left[\bra{\nu+1}\rho(t)\ket{\nu}-\bra{\nu}\rho(t)\ket{\nu+1}\right]\,,
\end{split}
\end{equation}
where $\nu\in\{A,B,C\}$, and $A=1$, $B=A+1$, $C=B+1$, and finally $C+1=A$. Note that $\phi(\nu,t)$ is obtained from $j(x,t)$ by discretizing the differentials. The total current in the quantum demon is then given by summing over all states and we have
\begin{equation}
\label{eq25}
\Phi(t)=\sum\limits_{\nu\in\{A,B,C\}} \phi(\nu,t)\,.
\end{equation}
In Fig.~\ref{phit} we plot the resulting total current $\Phi_\mrm{ss}(t)$ in the time-periodic stationary state for the examples considered earlier. 
As intuitively expected and in analogy to the classical model \cite{man12} there is a positive current, that is a current along $A\rightarrow B\rightarrow C\rightarrow A \rightarrow \dots$ driven by the interaction with the $N$ qubit stream, whose qubits are all prepared in the zero state \eqref{eq11}. However, we observe strong dependence on the coupling strength $\gamma$. Therefore, we will continue our analysis by considering the stationary current as a function of $\gamma$. For the sake of clarity it is instructive to introduce the mean stationary current, i.e. the current in the time-periodic state averaged over one interaction interval,
\begin{equation}
\label{eq26}
\overline{\Phi}=\frac{1}{\tau}\int_0^\tau\td t\,\Phi_\mrm{ss}(t)\,.
\end{equation}
In Figs.~\ref{phi_gamma} and \ref{phi_gamma2} we plot $\overline{\Phi} $ for a wide range of values of $\gamma$. In addition, we analyze how the current in the time-periodic stationary state depends on the initial preparation of the qubit stream. In Fig.~\ref{phi_gamma} we compare the outcome for the initial states,
\begin{subequations}
\begin{eqnarray}
\label{eq27}
\rho_n^1(0)&=&\begin{pmatrix} 1&0 \\0&0 \end{pmatrix}\\
\rho_n^2(0)&=&\begin{pmatrix} 0&0 \\0&1 \end{pmatrix}\\
\rho_n^3(0)&=&\begin{pmatrix} 1/2&0 \\0&1/2 \end{pmatrix}
\end{eqnarray}
\end{subequations}
and in Fig.~\ref{phi_gamma2} for the initial states
\begin{subequations}
\begin{eqnarray}
\rho_n^4(0)&=&\begin{pmatrix} 2/3&0 \\0&1/3 \end{pmatrix}\\
\rho_n^5(0)&=&\begin{pmatrix} 1/3&0 \\0&2/3 \end{pmatrix}\\
\rho_n^6(0)&=&\begin{pmatrix} 1/2&i/2 \\-i/2&1/2 \end{pmatrix}\label{eq28}\,.
\end{eqnarray}
\end{subequations}

In Fig.~\ref{phi_gamma} as well as in Fig.~\ref{phi_gamma2} we observe that the total current changes its sign if we inverse the initial state of the $N$ qubits. This behavior is in agreement with physical intuition and confirms the meaningfulness of the definition \eqref{eq24}. Further, we note that for equipartition, i.e. initial states $\rho_n^3(0)$ the current vanishes in the stationary state, which is illustrated by the green line falling on the $x$-axis in Fig~\ref{phi_gamma}. Furthermore, we observe that for initial states $\rho_n^1$, $\rho_n^2$, $\rho_n^4$, and $\rho_n^5$ the average current $\bar{\Phi}$ as a function of $\gamma$ has a pronounced maximum, which is a signature of a resonance effect between eigentime of the quantum demon, and the correlation time $1/\gamma$ describing the typical time scale of the interaction with the qubit stream.

\paragraph*{Fully quantum mechanical current}
Finally, with $\rho_n^6$ we analzye a ``truly'' quantum mechanical initial preparation of the qubits, which has \textit{no} classical equivalent. As illustrated in Fig.~\ref{phi_gamma2} we observe that the sign of the average current, $\bar{\Phi}$ strongly depends on the value of $\gamma$. This change of sign is a true quantum feature that originates in the ``quantum correlations'' of the initial preparation. Thus, such a behavior is not expected to be found in classical systems.

\paragraph*{Universal behavior for large $\gamma$}
All presented cases with initial states \eqref{eq27}-\eqref{eq28} have in common that for large values of $\gamma$ the total current vanishes. This can be understood by considering that the magnitude of the probability for transitions $\ket{A}\leftrightarrow \ket{C}$ is governed by $\gamma$. Hence for large $\gamma$ transitions $\ket{A}\leftrightarrow \ket{C}$ become much more likely than transitions $\ket{B}\leftrightarrow \ket{C}$ and $\ket{A}\leftrightarrow \ket{B}$, whose magnitude is of the order of one, cf. \eqref{eq03}. That means that for large $\gamma$ the quantum demon mostly oscillates between states $\ket{A}$ and $\ket{C}$ and almost never visits $\ket{B}$, i.e. no net current is driven through the quantum demon. Similar behavior can also expected in the classical, stochastic models studied in Refs.~\cite{man12,mandal_2013,strasberg_2013,barato_2013,barato_2013a}.

\subparagraph{Outlook -- possible extensions of the model}

\paragraph*{Thermal environments}
In the present analysis we restricted ourselves to the simplest possible model, namely a three state demon coupled to an information reservoir, here a $N$-qubit string. In traditional studies of Maxwell's demon one typically asks for processes that apparently violate the second law, and thus explore its validity. The present study is only a first step towards a ``full'' quantum version of the classic demon. Here, we merely showed that writing information can induce persistent currents, or in other word writing information can be used ``to do something''. Certainly, a next step would be to include thermal reservoirs into the model, in order to address more thermodynamic consequences. However, the description of open quantum systems would demand, for instance, to study dynamics generated by quantum master equations \cite{breuer_07}, which is beyond the scope of the present paper. Here we were focused on the simplest possible model, that shows demon-like behavior.

\paragraph*{Simplified description}
It has recently been shown by Barato and Seifert \cite{barato_2013a} that the description of the classical model of Ref.~\cite{man12} can be reduced to a two state model. Since in particular the treatment of a quantum demon coupled to thermal environments is much more involved that the classical analog, having such a reduced description would be very desirable. The focus of the present work, however, was to construct the quantum equivalent of the model proposed in \cite{man12}, and we leave possibly reduced descriptions for future study.

\subparagraph{Concluding remarks \label{sec:conclusion}}

In summary we have constructed a simple, solvable model of a quantum Maxwell demon. Our model represents the conceptual generalization of the minimal, classical model proposed in \cite{man12}. As a main result we have found that writing quantum information can induce persistent currents in stationary states, which is in complete analogy to the dynamical properties of minimal, classical models of Maxwell's demon. The present model obeys Hamiltonian dynamics, and no work or heat reservoirs have been included in the analysis. Therefore, the resulting probability current can not be used to perform thermodynamic work. For the time being we leave the extension of the model to include heat and work reservoirs for future study. The present analysis is intended to be a minimal and pedagogical illustration of a (quantum) information reservoir within the framework introduced in \cite{deffner_jarzynski_2013}. The current model offers a simple paradigm for investigating the quantum thermodynamic properties of information processing in a quantum Hamiltonian framework.

\acknowledgments{It is a pleasure to thank Christopher Jarzynski, Zhiyue Lu, and Dibyendu Mandal for stimulating discussions. We acknowledge financial support by a fellowship within the postdoc-program of the German Academic Exchange Service (DAAD, contract No D/11/40955) and from the National Science Foundation (USA) under grant DMR-1206971.}


\end{document}